\documentclass{PoS}
\usepackage{xcolor}

\title{Combined Search for Neutrinos from Dark Matter Annihilation in the Galactic Centre using ANTARES and IceCube}

\ShortTitle{Combined GC ANTARES IceCube DM Search}

\author{
The IceCube and ANTARES Collaborations\footnote{For collaborations list, see PoS(ICRC2019) 1177.}\\
{\itshape \href{http://icecube.wisc.edu/collaboration/authors/icrc19_icecube}{http://icecube.wisc.edu/collaboration/authors/icrc19\_icecube}}\\
{\itshape \href{http://antares.in2p3.fr/Collaboration/index2.html}{http://antares.in2p3.fr/Collaboration/index2.html}}\\
E-mail: \email{nadege.iovine@icecube.wisc.edu}\\
}

\abstract{
The ANTARES and IceCube neutrino telescopes have both searched for neutrinos from dark matter annihilation in the Galactic Centre, putting limits on the thermally-averaged dark matter self-annihilation cross section $\langle \sigma_A \upsilon \rangle$. For WIMP masses above 100 GeV, the most stringent limits were obtained by the ANTARES neutrino telescope, while for lower masses, limits achieved by IceCube are more competitive. The limits obtained by the two detectors are of comparable order of magnitude for WIMP masses going from 50 to 1000 GeV, making this mass range particularly interesting for a combined analysis. In this contribution, we present the limits of the first combined search for dark matter self-annihilation in the centre of the Milky Way using ANTARES and IceCube. The model parameters and the likelihood method were unified, thereby providing a benchmark for future dark matter searches conducted by each collaboration. By combining data of both detectors, we obtained improved limits with respect to the original limits published by the two collaborations.\\

\vspace{4mm}
{\bfseries Corresponding authors:}
\speaker{Nad\`ege Iovine}$^{1}$, Juan Antonio Aguilar Sanchez$^{1}$, Sebastian Baur$^{1}$, Sara Rebecca Gozzini$^{2}$, Juan de Dios Zornoza G\'omez$^{2}$\\
{$^{1}$ \itshape Universit\'e Libre de Bruxelles}\\
{$^{2}$ \itshape IFIC (UV-CSIC) - Instituto de F\'isica Corpuscular}
}

\FullConference{36th International Cosmic Ray Conference -ICRC2019-\\
		July 24th - August 1st, 2019\\
		Madison, WI, U.S.A.}

\begin{document}

\section{Introduction}\label{sec:intro}

Astrophysical observations strongly support the existence of dark matter. A prominent hypothesis is that dark matter is composed of cold non-baryonic particles which interact weakly with matter, the so-called Weakly Interacting Massive Particles (WIMPs). From observational evidence, we can deduce that galaxies are surrounded by a thermal relic dark matter halo, with a higher density towards the centre of the galaxy. This high concentration of dark matter in the centre of galaxies, such as our Milky Way, would favor the annihilation of WIMPs into secondary particles, specifically neutrinos. 

Indirect detection experiments, such as neutrino detectors, are looking for these secondary particles. The ANTARES and IceCube telescopes are already providing limits on the dark matter self-annihilation cross section \cite{ANTARES_GCWIMP, IC86_GCWIMP}. With this combined analysis, we aimed to enhance the indirect detection potential in the region where the two detectors are comparable, i.e.\@ from 50 to 1000 GeV. Another aspect was to address the differences between the two analysis methods and unify them, thus providing a benchmark for forthcoming dark matter analyses carried out by both experiments.

\section{Dark Matter}\label{sec:DM}
\begin{figure}[!tbp]
    \centering
    \begin{minipage}[b]{.49\textwidth}
    \includegraphics[width=\linewidth]{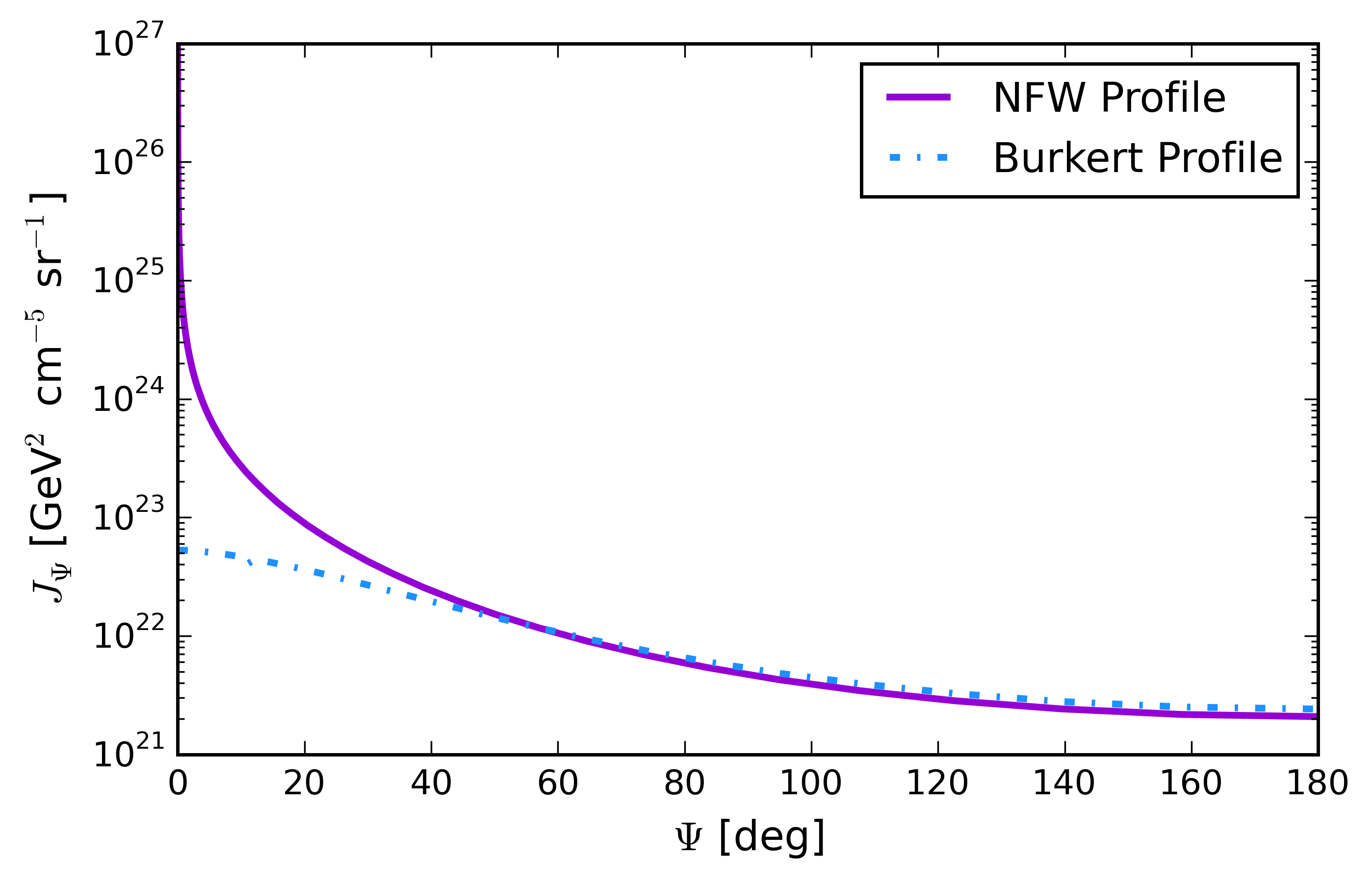}
    \end{minipage}
    \hfill
    \begin{minipage}[b]{.49\textwidth}
    \includegraphics[width=\linewidth]{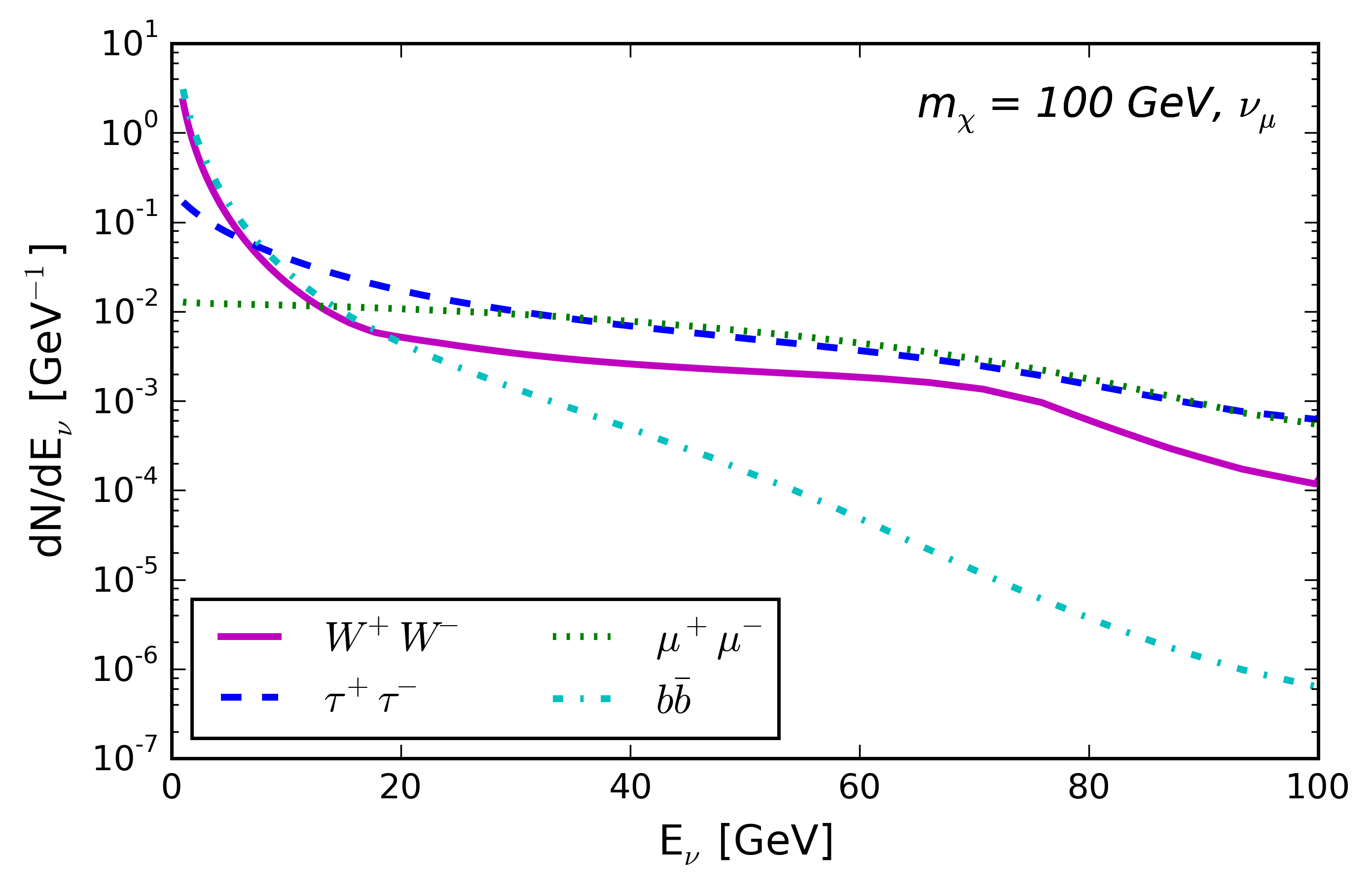}
    \end{minipage}
    \caption{\textbf{Left:} J-factors $J_{\Psi}$ as a function of the opening angle to the Galactic Centre $\Psi$ for the NFW and Burkert profiles. \textbf{Right:} Muon neutrino spectra at Earth for WIMP mass of 100 GeV and the four self-annihilation channels.}
    \label{fig:Model_parameters}
\end{figure}

The expected neutrino flux from dark matter annihilation in the Galactic Centre (GC) can be computed from the following equation:

\begin{equation}
    \frac{d\phi_{\nu}}{dE_{\nu}} = \frac{1}{2} \, \frac{\langle \sigma_A \upsilon \rangle}{4\pi m_{\chi}^2} \; \frac{dN_{\nu}}{dE_{\nu}} \int_{0}^{\Delta\Omega} \; d\Omega \int_{l.o.s} \rho_{\chi}^2(r) \, dl \, , 
    \label{eq:sig_expectation}
\end{equation}

\noindent where $m_{\chi}$ is the mass of the WIMP, $\langle \sigma_A \upsilon \rangle$ is the dark matter thermally-averaged self-annihilation cross section and $\frac{dN_{\nu}}{dE_{\nu}}$ is the differential number of neutrinos produced per annihilating WIMP pair. The integral term of the equation is referred to as the J-factor ($J_{\Psi}$) and is defined as the integral over the solid angle, $\Delta\Omega$, and line-of-sight (l.o.s) of the squared dark matter density $\rho_{\chi}$. For this work, two dark matter halo profiles were considered, the Navarro-Frenk-White (NFW) profile \cite{NFW} and the Burkert profile \cite{Burkert}:

\begin{equation}
    \rho_{NFW}(r) = \frac{\rho_0}{\frac{r}{r_s}\left( 1+ \frac{r}{r_s}\right)^2} \; , \; \; \; \; \; \rho_{Burkert}(r) = \frac{\rho_0}{\left(1+\frac{r}{r_s}\right) \left(1+ \left(\frac{r}{r_s}\right)^2\right)} \; ,
\end{equation}

\noindent where $r_s$ is the scale radius and $\rho_0$ is the characteristic dark matter density. These parameters were unified for both experiments and taken from \cite{Nesti}. Using these ingredients, we computed the J-factor as a function of the opening angle $\Psi$ using \textit{Clumpy} \cite{Clumpy}, as shown in the Figure \ref{fig:Model_parameters} (left).

In our effort to unify the two analysis methods, we noticed differences between the energy spectra used previously by both collaborations. While ANTARES was considering the spectra taken from \cite{Cirelli} known as PPPC4 tables, IceCube was using spectra computed with Pythia \cite{Pythia} for the IC86 GC Search. 
Furthermore, the spectra for dark matter annihilation to $\nu\bar{\nu}$ were approximated by a broadening of the neutrino line with a Gaussian for IceCube results, whereas the PPPC4 spectra includes additional electroweak corrections. For the purpose of our joint analysis, we decided to use the same spectra for both detectors, favouring the PPPC4 tables. A 100$\%$ branching ratio to $W^+W^-$, $\tau^+\tau^-$, $\mu^+\mu^-$ or $b\bar{b}$ is assumed. The neutrino spectra per annihilation process at Earth, $dN_{\nu}/dE_{\nu}$, for all annihilation channels considered and a WIMP mass of 100 GeV can be found in Figure \ref{fig:Model_parameters} (right).

\section{ANTARES and IceCube neutrino telescopes}\label{sec:detectors}

Due to the low neutrino interaction cross section, a large volume of material is required for neutrino detection. This is achieved by deploying a sparse array of photo-detectors in a deep, dark and dielectric medium, such as water or ice. These photo-sensors record the Cherenkov radiation induced by the secondary charged particles produced in the neutrino interactions in the surrounding medium.

\subsection{ANTARES}
ANTARES is a neutrino telescope located in the Mediterranean sea at coordinates 42$^{\circ}$48'N, 6$^{\circ}$10'E \cite{ANTARES}. This detector consists of an array of 885 optical modules attached to 12 strings with a length of 450 m. These strings are anchored to the seabed at a depth of about 2500 m and kept vertical by buoys. Each string holds 25 storeys composed of 3 optical modules (OMs) separated vertically by 14.5 m. The position of each string is monitored using a system of hydrophones and compasses installed within the detector volume.

\subsection{IceCube}
IceCube is a cubic-kilometer neutrino detector buried in the South Pole ice between depths of 1,450 m and 2,450 m \cite{IceCube}. This telescope consists of an array of 5,160 digital optical modules (DOMs) attached to 86 strings. In the centre of the detector, eight strings are deployed in a more compact way, forming the DeepCore subdetector. This denser configuration extends the detection of neutrinos to energies below 100 GeV.

\subsection{View of the Galactic Centre}
The main background of both experiments consists of atmospheric muons and neutrinos produced by the interaction of cosmic rays in the atmosphere. When considering up-going directions in the detectors, atmospheric muons are stopped by the Earth which considerably reduce the background. For IceCube, events with declinations corresponding to angles between $0^{\circ}$ to $90^{\circ}$ are seen as up-going events.  Since the Galactic Centre is located in the Southern Hemisphere ($\delta \sim - 29.01^{\circ}$), a veto is required to observe the centre of the galaxy. Unlike IceCube, ANTARES does not see the GC at a fixed zenith position in their local coordinates. As a result, declinations, $\delta$, below $-47^{\circ}$ are favoured in ANTARES since they are always below the horizon of the detector, while events with declinations between $-47^{\circ}$ and $47^{\circ}$ are seen as up-going only for part of the sidereal day. Thus, ANTARES has a privileged view of the GC with a visibility of $75\%$.

\section{Data sets}
For this analysis, we used data sets which were readily available in the ANTARES and IceCube collaborations. Both samples were optimized for the search of dark matter annihilation in the GC. Given the different scale and location of these experiments, different techniques were used to reduce the amount of atmospheric muon background for their respective event selections.

For ANTARES, the data sample considered is the same as for the dark matter search in the Milky Way using 9 years of data \cite{ANTARES_GCWIMP}. This data set is composed of track-like events recorded from 2007 to 2015, corresponding to a total of 2101.6 days. Two types of reconstructions are used depending on the deposited energy in the detector. The single-line reconstruction (QFit) is used to reconstruct events recorded on a single-line and gives the best fit for energies below 100 GeV. However, QFit only reconstructs the zenith direction of the event. The multi-line reconstruction ($\lambda$Fit) is used for higher energies and considers hits from several lines. The total number of events in the sample is 15651 for the $\lambda$Fit and 1077 for the QFit reconstruction.

For IceCube, the data sample of the IC86 GC WIMP search analysis \cite{IC86_GCWIMP} is used. This data selection consists of track-like events taken from the 15th of May 2012 to the 18th of May 2015 with the 86-string configuration, for a total of 1007 days. The sample considered is composed of a total of 22622 events. The event selection was optimized for muon neutrinos, but all neutrino flavors are considered in the signal simulation samples.

\section{Analysis Method}\label{sec:Analysis}
In order to search for an excess of signal neutrinos from the Galactic Centre by combining data from both ANTARES and IceCube neutrino telescopes, a binned likelihood method is used.

\subsection{Likelihood Method}

The binned likelihood is defined as the product over all bins of the Poisson probability to observe $n_{obs}(i)$ in a bin i:
\begin{equation}
    \mathcal{L}(\mu) = \prod_{i=min}^{max} Poisson\left(n_{obs}(i) \, ; \, n_{obs}^{tot} \, f(i; \mu)\right) \, , 
\end{equation}

\noindent where $\mu \in [0,1]$ corresponds to the signal fraction of the total sample and is the maximisation variable. With this method, the observed number of events in a given bin i, $n_{obs}(i)$, is compared to the number of expected events, which is determined from the total number of data events in the histogram, $n_{obs}^{tot}$, and the total fraction of events within the specific bin:
\begin{equation}
    f(i \, ;  \, \mu) = \mu \, f_s(i) \, + \, (1 - \mu) \, f_{BG}(i) \, ,
\end{equation}
 
 \noindent with $f_s$ and $f_{BG}$ being respectively the signal and the background probability density functions (PDFs). This method is used to get the best estimate on the signal fraction, $\widehat{\mu}$, which can be obtained by maximising the likelihood, $\mathcal{L}(\mu)$. If this value is consistent with the background only hypothesis, the upper limit on the signal fraction, $\mu_{90\%}$, is computed using the Feldman-Cousins method \cite{Feldman-cousins}. The limit on $\langle\sigma_A\upsilon\rangle$ can then be deduced from $\mu_{90\%}$ using the expected number of signal neutrinos, obtained from equation \ref{eq:sig_expectation}, and the total number of observed events.

\subsection{Probability Density Function}

\begin{figure}[!tbp]
    \centering
    \begin{minipage}[b]{.8\textwidth}
    \includegraphics[width=\linewidth]{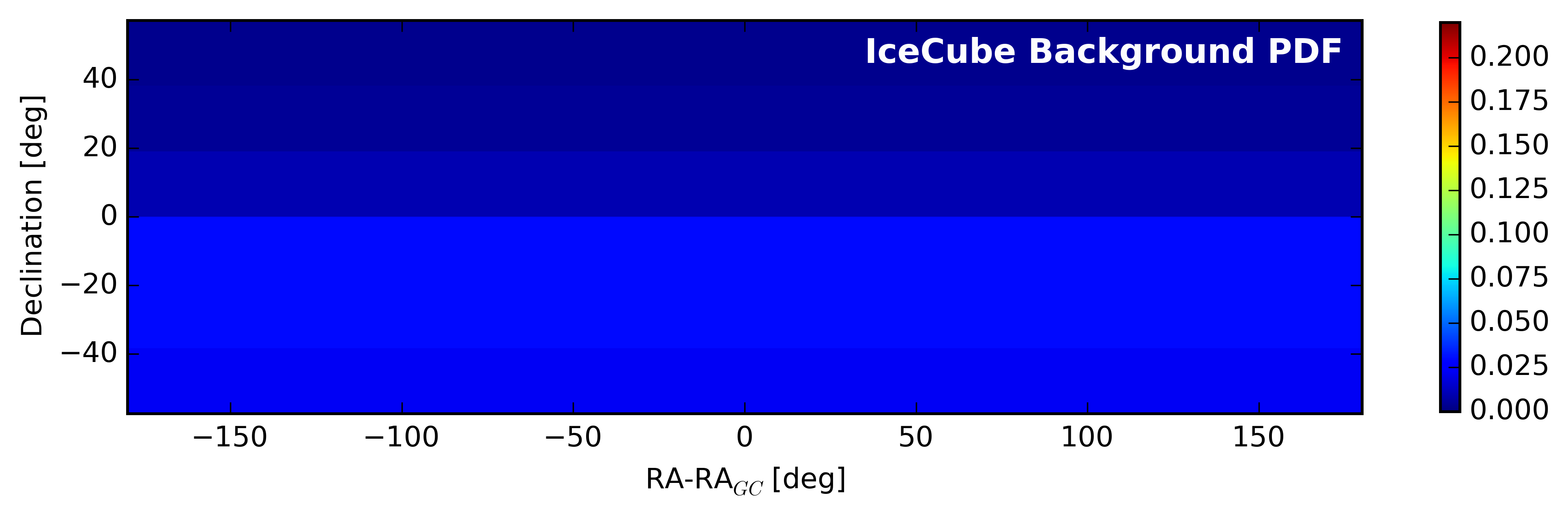}
    \end{minipage}
    \hfill
    \begin{minipage}[b]{.8\textwidth}
    \includegraphics[width=\linewidth]{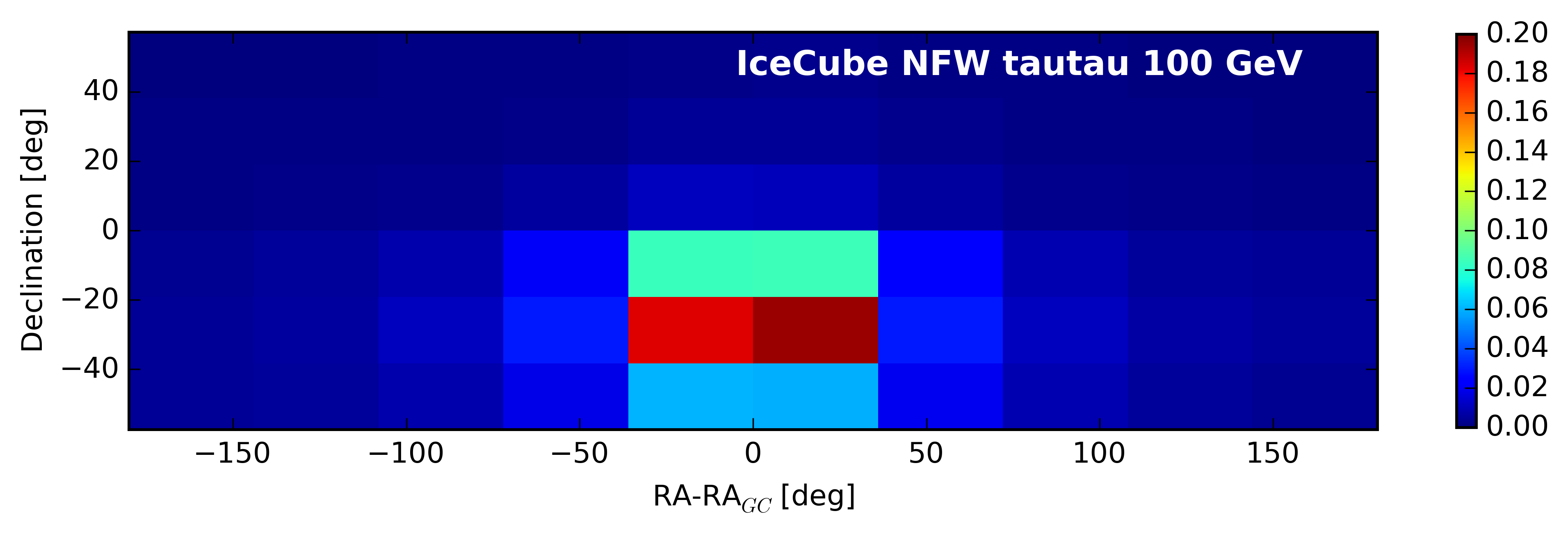}
    \end{minipage}
    \caption{\textbf{Top:} IceCube background PDF obtained from data scrambled in RA, where the color scale express the density. \textbf{Bottom:} IceCube signal PDF $\tau^+\tau^-$ channel and $m_{\chi}$ = 100 GeV assuming the NFW profile.}
    \label{fig:IceCube_PDFs}
\end{figure}

For IceCube, we are considering PDFs composed of 2-dimensional distributions of the events in right ascension (RA) and declination (dec). The binning consists of 10 bins in RA ranging from -$2\pi$ to $2\pi$ and 6 bins in declination covering a range of -1 to 1 rad (see Figure \ref{fig:IceCube_PDFs}). Since the background is uniform in RA, the background PDF is obtained by scrambling data in RA.

\begin{figure}[!tbp]
    \centering
    \begin{minipage}[b]{.49\textwidth}
    \includegraphics[width=\linewidth]{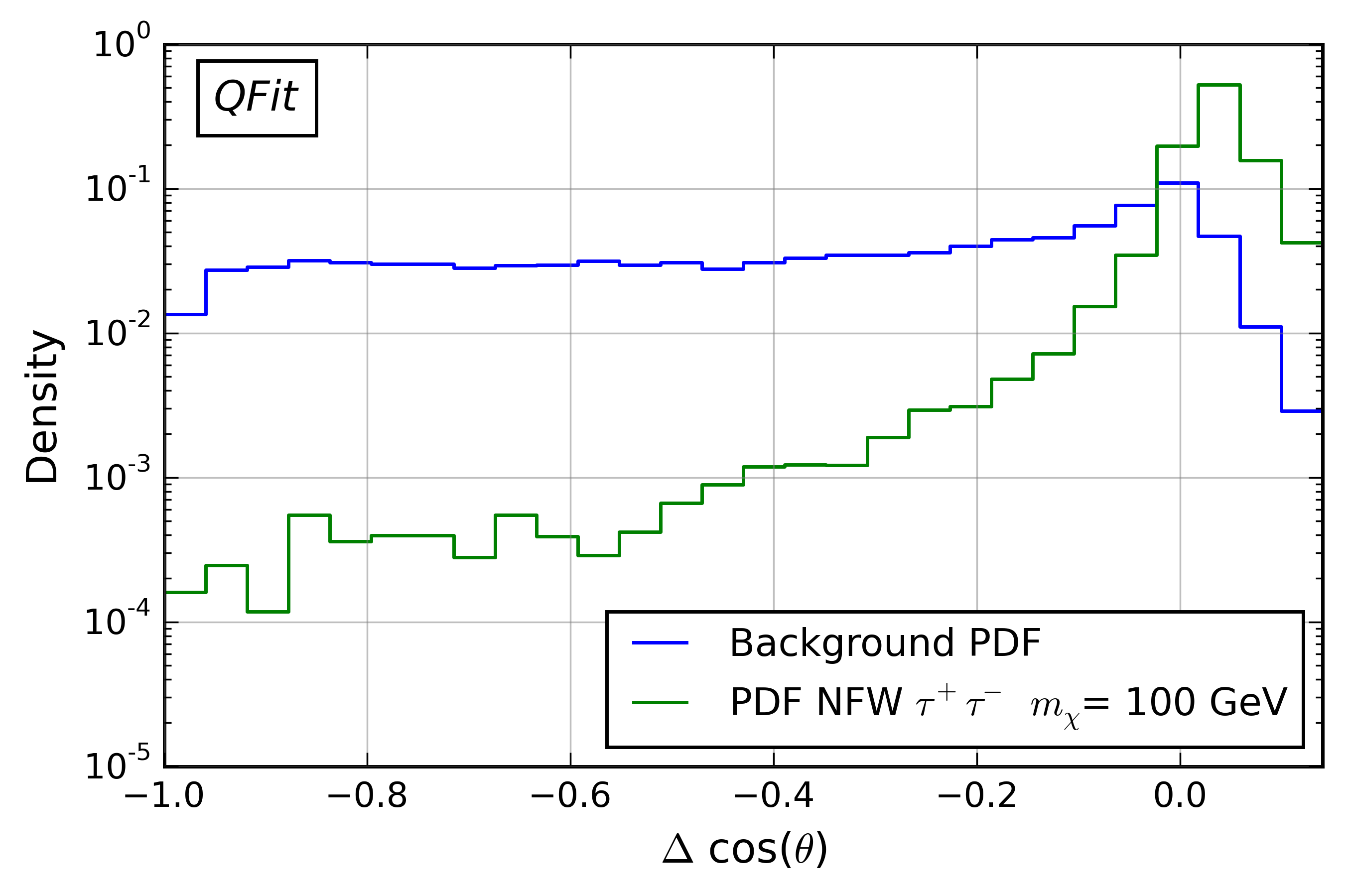}
    \end{minipage}
    \hfill
    \begin{minipage}[b]{.49\textwidth}
    \includegraphics[width=\linewidth]{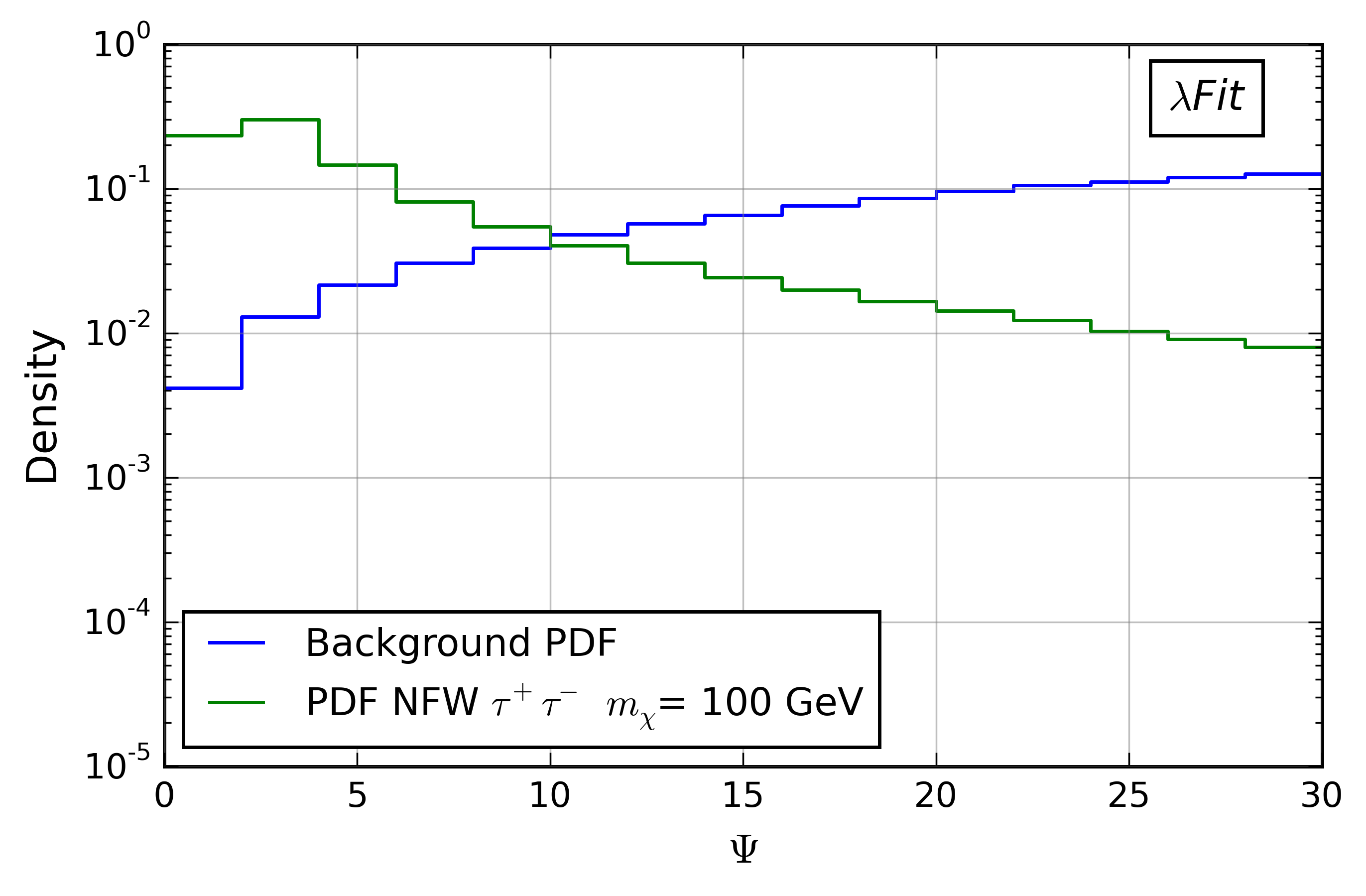}
    \end{minipage}
    \caption{ANTARES background and signal PDFs for the $\tau^+\tau^-$ channel and NFW profile with $m_{\chi}$ = 100 GeV for both QFit (left) and $\lambda$Fit (right).}
    \label{fig:ANTARES_PDFs}
\end{figure}

In the case of ANTARES, 1-dimensional distributions of the events are used, which are different for the two reconstructions. For QFit, the histograms are composed of 28 bins in $\Delta \cos(\theta)$ ranging from -1 to 0.14 rad, where $\Delta \cos(\theta) = \cos(\theta_{GC}) - \cos(\theta_{event})$ (see Figure \ref{fig:ANTARES_PDFs} left), $\theta_{event}$ is the zenith of the event and $\theta_{GC}$ is the zenith of the GC. For $\lambda$Fit, the events are distributed in 15 bins in $\Psi$ distributed from 0$^{\circ}$ to 30$^{\circ}$, where $\Psi$ is the angular difference between the event and the Galactic Centre (see Figure \ref{fig:ANTARES_PDFs} right).

\subsection{Combined Likelihood}
Once computed for ANTARES and IceCube separately, the likelihoods are combined in a single likelihood defined as:


\begin{equation}
    \mathcal{L}_{comb}(\mu) = \prod^{A,I}_{k} \, \mathcal{L}_{k}(\mu_k) \, ,
\end{equation}

\noindent where the indexes A and I are allocated to the ANTARES and IceCube likelihoods, respectively. As a result, the only parameter to maximise is the joint signal to background ratio, $\mu$, which can be expressed as:

\begin{equation}
    \mu = \frac{N_{sig}}{N_{tot}
    } = \frac{N_{sig}^{A} + N_{sig}^{I}}{N_{tot}^A + N_{tot}^I} =  \frac{N_{sig} (s_{A}+s_{I})}{N_{tot}(b_{A}+b_{I})} \, ,
\end{equation}

\noindent where $N_{sig}$ is the total number of expected signal events, obtained by summing the individual number of signal events from the two experiments $N_{sig}^{A}$ and $N_{sig}^{I}$. Likewise, the total number of observed events $N_{tot}$ is the sum of $N_{tot}^{A}$ and $N_{tot}^{I}$. We define the relative signal efficiencies of each sample, $s_k$, as the ratio $N_{sig}^{k}/N_{sig}$ and the relative background efficiencies as $N_{tot}^{k}/N_{tot}$.
We can then write the individual signal fractions as:

\begin{equation}
    \mu_k = \frac{N_{sig}^{k}}{N_{tot}^k} = \frac{s_k \, N_{sig}}{b_k \, N_{tot}} = w_k \frac{N_{sig}}{N_{tot}} \, . 
 \end{equation}

\noindent The weight associated with each experiment, $w_k$, is then used to quantify the contribution of each sample to the total likelihood.

\section{Results}\label{sec:Results}
\begin{figure}[!tbp]
    \centering
    \begin{minipage}[b]{.49\textwidth}
    \includegraphics[width=\linewidth]{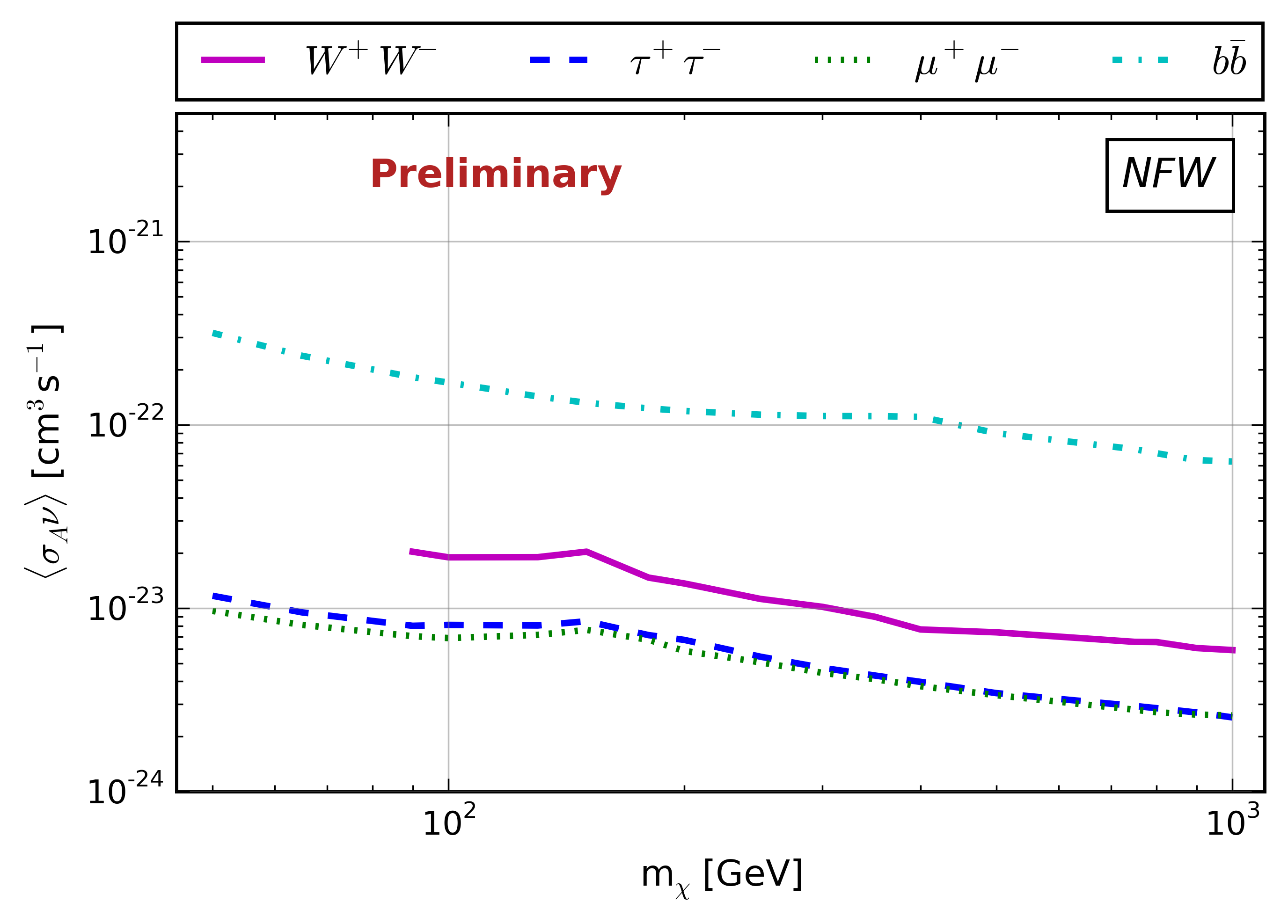}
    \end{minipage}
    \hfill
    \begin{minipage}[b]{.49\textwidth}
    \includegraphics[width=\linewidth]{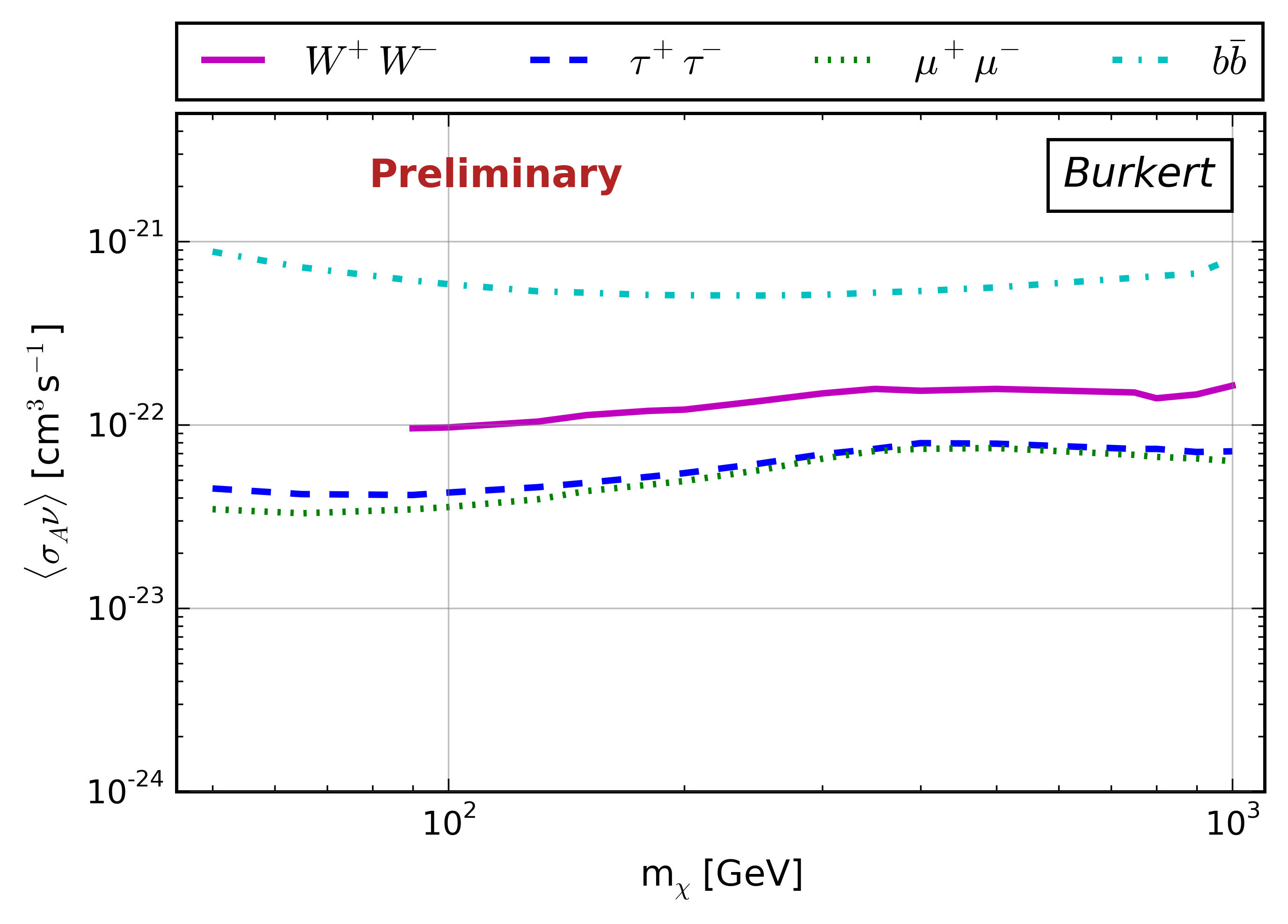}
    \end{minipage}
    \caption{Final combined limits on the thermally-averaged dark matter annihilation cross section $\langle\sigma_A\upsilon\rangle$ as a function of the WIMP mass $m_{\chi}$. All annihilation channels considered for this analysis are presented ($W^+W^-$, $\tau^+\tau^-$, $\mu^+\mu^-$,  $b\bar{b}$) for both the NFW (left) and Burkert (right) halo profiles.}
    \label{fig:limits_all}
\end{figure}

For this joint analysis, we made use of data collected by the ANTARES and IceCube neutrino telescopes during a period of 9 years and 3 years, respectively. By combining these data samples on the likelihood level, we observed no significant neutrino excess over the background in the direction of the Galactic Centre. Therefore, we present here the combined limits on the thermally-averaged dark matter self-annihilation cross section $\langle\sigma_A \upsilon\rangle$. In Figure \ref{fig:limits_all}, the 90$\%$ upper limits for all annihilation channels are shown separately for the NFW (left) and Burkert (right) halo profiles. 

\begin{figure}[!tbp]
    \centering
    \begin{minipage}[b]{.49\textwidth}
    \includegraphics[width=\linewidth]{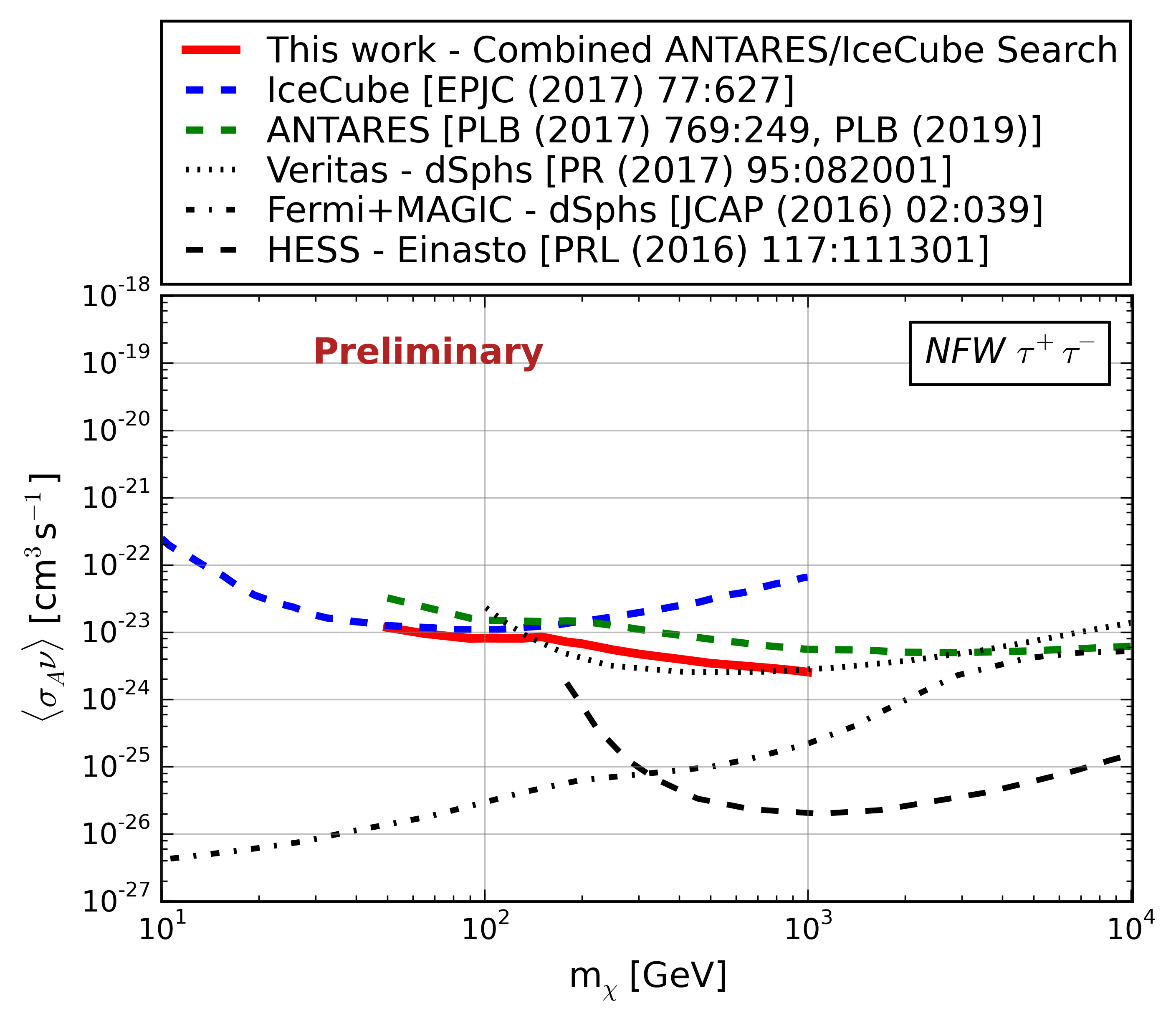}
    \end{minipage}
    \hfill
    \begin{minipage}[b]{.49\textwidth}
    \includegraphics[width=\linewidth]{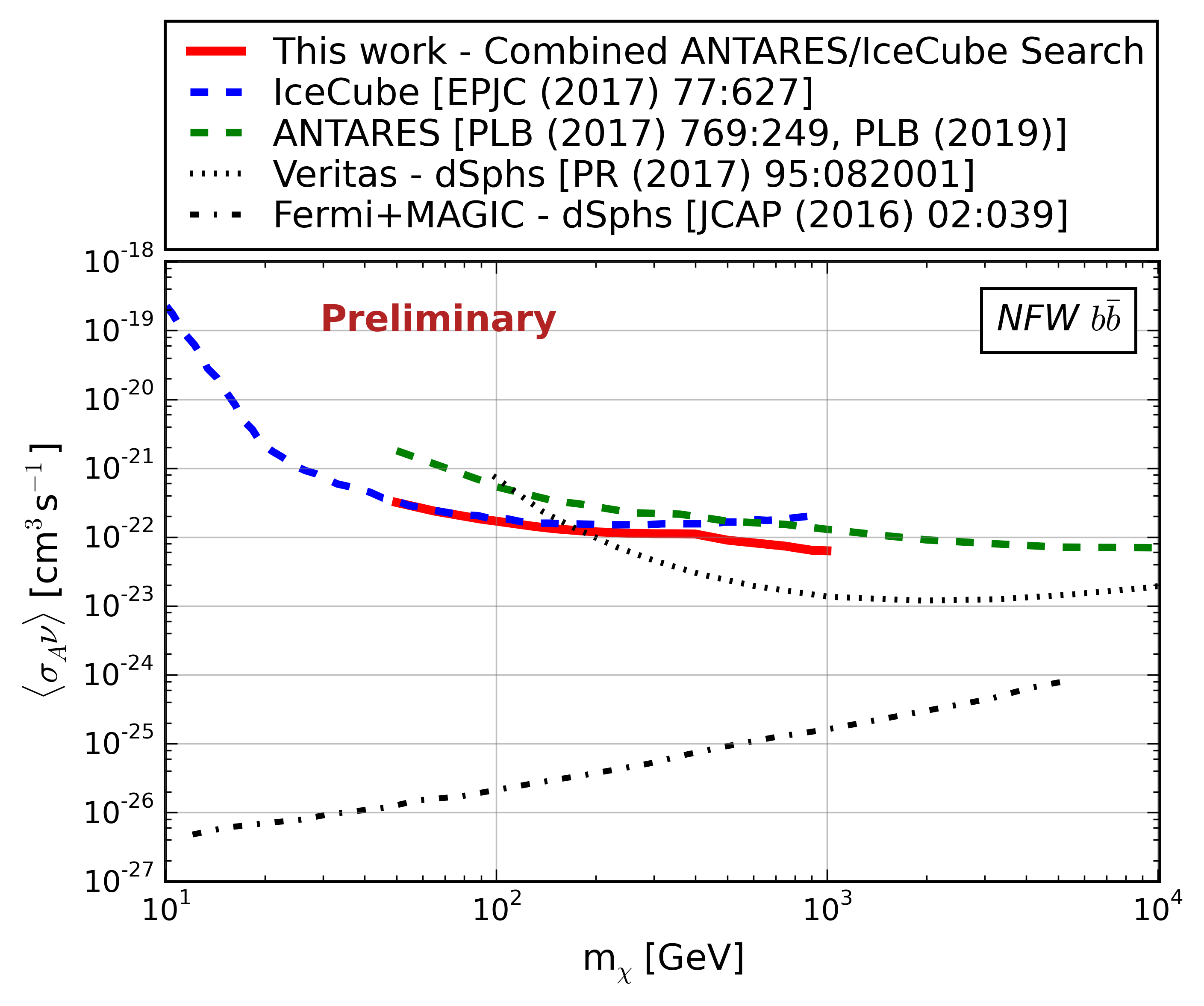}
    \end{minipage}
    \caption{Limit on the thermally-averaged dark matter annihilation cross section $\langle\sigma_A\upsilon\rangle$ obtained for the combined analysis as a function of the WIMP mass $m_{\chi}$ assuming the NFW halo profile for the $\tau^+\tau^-$ (left) and $b\bar{b}$ (right) annihilation channel. Also shown are limits from IceCube \cite{IC86_GCWIMP}, ANTARES \cite{ANTARES_GCWIMP}, VERITAS \cite{Veritas}, Fermi-MAGIC \cite{Fermi-MAGIC}, as well as HESS \cite{HESS}. It should be noted that this combined analysis treated under-fluctuations differently than what is presented in the ANTARES paper. When obtaining limits with lower values than sensitivities, this joint analysis kept the limit as obtained from the likelihood method while the limits were moved to sensitivities for ANTARES.}
    \label{fig:limits_ comparison}
\end{figure}

By way of comparison, the combined limits are shown in Figure \ref{fig:limits_ comparison} alongside the ANTARES and IceCube limits obtained previously by each experiment. These limits were computed with the same data sets used for this analysis. Both $\tau^+\tau^-$ (left) and $b\bar{b}$ (right) annihilation channels are represented assuming the NFW halo profile. When compared to the individual IceCube and ANTARES limits, the combined limits show improvements for the WIMP mass range considered, i.e.\@ from 50 GeV to 1 TeV. 
An enhancement can be seen for almost all annihilation channels and the two DM halo profiles considered. Only the $b\bar{b}$ channel combined limit obtained for the Burkert profile is dominated by IceCube, which has a better limit than ANTARES for the entire mass range for that particular case.

Although this analysis demonstrates what we can gain from a combined DM search, it can still be improved. First, the data sets considered here were not specifically designed for this analysis and were thus restricting the choice of some parameters, such as the WIMP mass range used. While it was decided to use data sets which already led to publications for this work, the analysis could be extended to more years of data in the future. The likelihood method could also be revised by moving to an unbinned likelihood.

\bibliographystyle{ICRC}
\bibliography{references}

\end{document}